\documentstyle[aps,epsf]{revtex}

\begin{document}

\title{Quintessence and Cosmic Acceleration}
\author{M. K. Mak\footnote{E-mail:mkmak@vtc.edu.hk}}
\address{Department of Physics, The Hong Kong University of Science and Technology,Clear Water Bay, Hong Kong, P. R. China.}
\author{T. Harko\footnote{E-mail: tcharko@hkusua.hku.hk}}
\address{Department of Physics, The University of Hong Kong,
Pokfulam Road, Hong Kong, P. R. China.}
\maketitle

\begin{abstract}

A cosmological model with perfect fluid and self-interacting quintessence field is considered
in the framework of the spatially flat Friedmann-Robertson-Walker (FRW) geometry. By assuming that all physical
quantities depend on the volume scale factor of the Universe, the general solution of the
gravitational field equations can be expressed in an exact parametric form. The
quintessence field is a free parameter. With an appropriate choice of the scalar field a class of exact solutions
is obtained, with an exponential type scalar field potential fixed via the gravitational field equations.
The general physical behavior of the model is consistent with the recent cosmological scenario favored by supernova type Ia
observations, indicating an accelerated expansion of the Universe.

\end{abstract}

\section{Introduction}

Measurements of the cosmic microwave background, the mass power
spectrum and of the luminosity-red shift relation observed for Type Ia
supernovae with redshift up to about $z\lesssim 1$ provided evidence that we
may live in a low mass-density Universe, with the contribution of the
non-relativistic matter (baryonic plus dark) to the total energy density of
the Universe of the order of $\ \Omega _{m}\sim 0.3$ \cite{1}-\cite{3}. The value of $%
\Omega _{m}$ is significantly less than unity \cite{4} and consequently either
the Universe is open or there is some additional energy density $\rho $
sufficient to reach the value $\Omega _{total}=1$, predicted by inflationary
theory. The missing energy should possess negative pressure $p$ and equation
of state $\gamma =\frac{p}{\rho }<0$. Observations also show that the
deceleration parameter of the Universe $q$ is in the range $-1\leq q<0$, and
the present-day Universe undergoes an accelerated expansionary evolution.

Several physical models have been proposed to give a consistent physical
interpretation to these observational facts. One candidate for the missing
energy is vacuum energy density or cosmological constant $\Lambda $,
satisfying the equation of state $p=-\rho $ \cite{5}. According to this
cosmological model the present-day Universe consists of a mixture of vacuum
energy and cold dark matter. The conditions under which the dynamics of a
self-interacting Brans-Dicke (BD) field can account for the accelerated
expansion of the Universe have been considered in \cite{6}. In this scenario
accelerated expanding solutions can be obtained with a quadratic
self-coupling of the BD field and a negative coupling constant $\omega $.
But a cosmic fluid (pressureless and with pressure) obeying a perfect fluid
type equation of state cannot support the acceleration in the BD theory.

Another possibilities are cosmologies based on a mixture of cold dark matter
and quintessence, a slowly-varying, spatially inhomogeneous component \cite{7}.
An example of implementation of the idea of quintessence is the suggestion
that it is the energy associated with a scalar field $\phi $ with
self-interaction potential $U(\phi )$. If the potential energy density is
greater than the kinetic one, then the pressure $p_{\phi }=\frac{1}{2}\dot{%
\phi}^{2}-U(\phi )$ associated to the $\phi $-field is negative.
Quintessencial cosmological models have been recently intensively
investigated in the physical literature \cite{8}.
For quintessence, the equation of state defined by $\gamma _{\phi }=\frac{%
p_{\phi }}{\rho _{\phi }}$ is subject to the constraint $-1\leq \gamma
_{\phi }\leq -0.6$.

But all these scenarios present two major difficulties \cite{9}. The first one is
the fine-tunning problem: why is the missing energy density today so small
compared to typical particle physics scales? If today $\ \Omega _{m}\sim 0.3$%
, then the missing energy density is of the order of $10^{-47}GeV$, which
appears to require the introduction of a new mass scale of $14$ orders of
magnitude smaller than the electroweak scale. The second problem is the
''cosmic coincidence'' problem: since the missing energy density and the
matter density decrease at different rates as a result of the expansion of
the Universe, their ratio must be set to a specific, infinitesimal value in the very early Universe
in order for the two densities to nearly coincide  
today. In order to solve these problems a form of the quintessence field,
called ''tracker fields'', has been introduced, in which the tracker field $%
\phi $ rolls down the potential $U(\phi )$ according to an attractor-like
solution to the equations of motion \cite{9}. 

Most of the studied quintessence models involve minimally coupled scalar
fields with different potentials. One of the must studied is the simple exponential
potential. But in this case the energy density is not enough to close to 
$\Omega =1$ the total energy density of the Universe. To solve the cosmic
quintessence problem inverse power law potentials have been considered
\cite{10}. For this potential the predicted value of $\gamma _{\phi }$ \ is not
in good agreement with observations. The implications for the cosmological
evolution of potentials, which  asymptotically behaves  like the inverse
power law or exponential potentials, of the form $U_{0}\left[ \cosh \left(
\lambda \phi \right) -1\right] ^{p}$ \cite{11} or
$U_{0}\left[ \sinh \left( \alpha \sqrt{k_{0}}\Delta \phi \right) \right]
^{\beta }$ \cite{12}, have also been studied.

The nature of the scalar field potential compatible with a power law
expansion in a self-interacting BD cosmology with a perfect fluid background
has been analyzed in \cite{13}. The form of the cosmic potential has been reconstructed for a minimally coupled quintessence field,
in \cite{14}, by using observational data, from the
expression of the luminosity distance $d_{L}(z)$ as function of the redshift $z$.
General cosmological models with non-minimal
coupling of the scalar field have been considered in \cite{15}.

It is the purpose of the present Letter to consider some exact classes of
solutions of the gravitational field equations for a spatially
flat FRW Universe composed from a mixture of matter and quintessence field. By
assuming that all physical quantities depend on the volume scale factor
of the Universe only, the general solution of the field equations can be
expressed in an exact parametric form, with the volume scale factor taken as
parameter and the quintessence field $\phi $ as a free parameter. By
appropriately choosing the mathematical form of $\phi $ a cosmological model
is obtained which incorporates the basic observational properties of the
present day Universe. 

The present Letter is organized as follows. The basic equations of the model
and the general solutions are obtained in Section II. The general solution to the field equations
corresponding to dust matter is presented in Section III. In Section IV we discuss and conclude our
results.

\section{Classes of general solutions of field equations with matter and quintessence}

We assume that the matter content of the Universe is composed of a perfect
fluid with the pressure $p_{m}$ and  density $\rho _{m}$,
obeying a barotropic equation of state of the form $p_{m}=\left( \gamma
-1\right) \rho _{m},1\leq \gamma \leq 2$ and a scalar field $\phi $ as the source of the
quintessence matter, with pressure $p_{\phi }=\frac{1}{2}\dot{\phi}%
^{2}-U\left( \phi \right) $ and energy density $\rho _{\phi }=\frac{1}{2}\dot{\phi}%
^{2}+U\left( \phi \right) $. $U\left( \phi \right) $ is the potential
energy of the quintessence field. 

The line element for a spatially flat FRW
cosmological model takes the form
\begin{equation}\label{1}
ds^{2}=dt^{2}-a^{2}\left( t\right) \left( dx^{2}+dy^{2}+dz^{2}\right).
\end{equation}

In the present paper we use units so that $8\pi G=c=1$.

By introducing the Hubble function $H=\frac{\dot{a}}{a}=\frac{\dot{V}}{3V}$
(where a \ dot $\cdot =\frac{d}{dt})$ and $V(t)=\left( \frac{a(t)}{a_{0}}\right) ^{3}$, $a_{0}=constant>0$, the gravitational
field equations and the equation of motion of the quintessence field are:
\begin{equation}\label{2}
3H^{2}=\rho _{m}+\rho _{\phi },2\dot{H}+3H^{2}=-\left( \gamma -1\right) \rho
_{m}-p_{\phi },\ddot{\phi}+3H\dot{\phi}=-\frac{dU\left( \phi \right) }{d\phi }.    
\end{equation}

The energy conservation equation of the baryonic matter, which follows from the Bianchi identity
gives
$\dot{\rho}_{m}+3H\gamma \rho _{m}=0$.  

Therefore the time evolution of the energy density of the 
matter component of the Universe, obeying a barotropic equation of state, is given by
$\rho _{m}=\rho _{0}V^{-\gamma }$,  
where $\rho _{0}$ is a positive constant of integration.

The observationally important physical quantity, the deceleration parameter $%
q$ is defined according to $q=\frac{dH^{-1}}{dt}-1$. The sign of the deceleration
parameter indicates whether the cosmological model accelerates or
decelerates. The positive sign corresponds to decelerating models whereas
the negative sign indicates accelerated expansion.

By substracting Eqs. (\ref{2}) we obtain
\begin{equation}\label{3}
2\dot{H}+\dot{\phi}^{2}+\gamma \rho _{m}=0.  
\end{equation}

With the use of the definition of the Hubble function and of Eqs. (\ref{2}), Eq. (\ref{3}) can be transformed to the form
\begin{equation}\label{4}
\frac{\ddot{V}}{V}-\left( \frac{\dot{V}}{V}\right) ^{2}+\frac{3}{2}\gamma
\rho _{0}V^{-\gamma }+\frac{3}{2}\dot{\phi}^{2}=0.  
\end{equation}

By means of the substitution $\dot{V}=u$ and $\ddot{V}=u\frac{du}{dV}=\frac{1}{2}\frac{d}{dV}u^{2}$
we obtain the following first
order linear differential equation in the variable $u^{2}$ describing the dynamics of the Universe 
for a mixture of matter and scalar quintessence field:
\begin{equation}\label{8}
\frac{d\left( u^{2}\right) }{dV}+\left[ 3V\left( \frac{d\phi }{dV}\right)
^{2}-\frac{2}{V}\right] u^{2}=-3\gamma \rho _{0}V^{1-\gamma }.  
\end{equation}

On integration, we obtain the solution of Eq. (\ref{8}) in the form
\begin{equation}\label{9}
u^{2}=V^{2}e^{-3\int V\left( \frac{d\phi }{dV}\right) ^{2}dV}\left[
g-3\gamma \rho _{0}\int V^{-1-\gamma }e^{3\int V\left( \frac{d\phi }{dV}%
\right) ^{2}dV}dV\right],   
\end{equation}
where $g$ is a constant of integration. On further integration, we obtain the general solution of Eq. (\ref{4}): 
\begin{equation}\label{10}
t-t_{0}=\int \frac{e^{\frac{3}{2}\int V\left( \frac{d\phi }{dV}\right)
^{2}dV}}{V\sqrt{g-3\gamma \rho _{0}\int V^{-1-\gamma }e^{3\int V\left( \frac{%
d\phi }{dV}\right) ^{2}dV}dV}}dV. 
\end{equation}
where $t_{0}$ is a constant of integration.

Therefore the general solution of the Einstein's
gravitational field equations with matter and quintessence field can be expressed in the following exact
parametric form, with $V>0$ taken as parameter:
\begin{equation}\label{11}
\phi =\phi \left( V\right) ,a=a_{0}V^{\frac{1}{3}},\rho _{m}=\rho
_{0}V^{-\gamma },p_{m}=\left( \gamma -1\right) \rho _{m},  
\end{equation}
\begin{equation}\label{12}
U\left( V\right) =\frac{V^{2}\left[ g-3\gamma \rho _{0}\int V^{-1-\gamma
}e^{3\int V\left( \frac{d\phi }{dV}\right) ^{2}dV}dV\right] \left[ \frac{1}{3%
}V^{-2}-\frac{1}{2}\left( \frac{d\phi }{dV}\right) ^{2}\right] }{e^{3\int
V\left( \frac{d\phi }{dV}\right) ^{2}dV}}-\rho _{0}V^{-\gamma },  
\end{equation}
\begin{equation}\label{13}
\rho _{\phi }\left( V\right) =\frac{1}{3}\frac{\left[ g-3\gamma \rho
_{0}\int V^{-1-\gamma }e^{3\int V\left( \frac{d\phi }{dV}\right) ^{2}dV}dV%
\right] }{e^{3\int V\left( \frac{d\phi }{dV}\right) ^{2}dV}}-\rho
_{0}V^{-\gamma },  
\end{equation}
\begin{equation}\label{14}
p_{\phi }\left( V\right) =\frac{V^{2}\left[ g-3\gamma \rho _{0}\int
V^{-1-\gamma }e^{3\int V\left( \frac{d\phi }{dV}\right) ^{2}dV}dV\right] %
\left[ \left( \frac{d\phi }{dV}\right) ^{2}-\frac{1}{3}V^{-2}\right] }{%
e^{3\int V\left( \frac{d\phi }{dV}\right) ^{2}dV}}+\rho _{0}V^{-\gamma },
\end{equation}
\begin{equation}\label{15}
q\left( V\right) =\frac{9}{2}\left\{ V^{2}\left( \frac{d\phi }{dV}\right)
^{2}+\frac{\gamma \rho _{0}e^{3\int V\left( \frac{d\phi }{dV}\right) ^{2}dV}%
}{V^{\gamma }\left[ g-3\gamma \rho _{0}\int V^{-1-\gamma }e^{3\int V\left( 
\frac{d\phi }{dV}\right) ^{2}dV}dV\right] }\right\} -1.  
\end{equation}

Classes of exact solutions of the field equations can be easily generated by the
appropriate choice of the scalar field $\phi $ as an arbitray function of
volume scale factor $V$. Via the gravitational field equations the choice of the
quintessence field also fixes the mathematical form of the potential.

\section{Cosmological dust solution with  quintessence field}

Assuming that at the present period the Universe is cold baryonic matter
dominated with negligible thermodynamic pressure, we can put in the field equations $p_{m}=0$,
or, equivalently, we choose $\gamma =1$. 

In order to have a clear physical interpretation of the
solutions, we choose the scalar field $\phi $ in the form
$\phi =\sqrt{n}\ln V$,  
where $n$ is a positive constant.

With this choice of the scalar field $\phi $ we obtain a solution of the Einstein gravitational field equations with dust matter and quintessence, which can be expressed in the
following exact parametric form:
\begin{equation}\label{16}
t-t_{0}=\int \frac{V^{\frac{3n}{2}-1}}{\sqrt{g+\frac{3\rho _{0}}{1-3n}%
V^{3n-1}}}dV,a=a_{0}e^{\frac{\phi }{3\sqrt{n}}}, \rho _{m}=\frac{\rho _{0}}{V},  
\end{equation}
\begin{equation}\label{18}
U\left( V\right) =\left( \frac{1}{3}-\frac{n}{2}\right) F\left( V\right)
-\rho _{0}V^{-1}, \rho _{\phi }\left( V\right) =\frac{1}{3}F\left( V\right) -\rho _{0}V^{-1}, 
\end{equation}
\begin{equation}\label{20}
p_{\phi }\left( V\right) =\left( n-\frac{1}{3}\right) F\left( V\right) +\rho
_{0}V^{-1}, q\left( V\right) =\frac{9}{2}\left[ n+\frac{\rho _{0}}{VF\left( V\right) }%
\right] -1,  
\end{equation}
where $F(V)=gV^{-3n}+\frac{3\rho _{0}}{1-3n}V^{-1}$.

The time transformation can be evaluated and expressed in terms of hypergeometric
functions. 

In order that the Universe undergoes an accelerating phase, it is
necessary that $q<0$. By imposing this condition and with the use of Eqs. (\ref{18}) and (\ref{20}), we obtain the following constraint on the parameter $n$:
\begin{equation}
n<\frac{2\frac{\rho _{\phi }}{\rho _{m}}-1}{9\left( \frac{\rho _{\phi }}{%
\rho _{m}}+1\right) }.
\end{equation}

By assuming a ratio of the quintessence field energy density and matter
energy density of the order of
$\frac{\rho _{\phi }}{\rho _{m}}\approx \frac{7}{3}$ [1-3], we find for the parameter $n$ the constraint $n<0.12$. By assuming
for the ratio $\frac{\rho _{\phi }}{\rho _{m}}$ a value of $4$ ($\rho _{\phi }\propto 0.8$, $\rho _{m}\propto 0.2$), we obtain for $n$
the condition $n<0.155$. 

The form of the quintessence field potential follows from Eqs. (\ref{16}) and (\ref{18})
and is given, as a function of the field by
\begin{equation}
U\left( \phi \right) =3\rho _{0}\frac{n}{2\left( 1-3n\right) }e^{-\frac{\phi 
}{\sqrt{n}}}+g\left( \frac{1}{3}-\frac{n}{2}\right) e^{-3\sqrt{n}\phi }.
\end{equation}

The dependence of the quintessence potential-quintessence field is of
exponential type.

\section{Discussions and Final Remarks}

In the present paper we have presented the general solution of the gravitational field equations in a flat geometry
for a mixture of matter and quintessence field, with the field $\phi $ as a free function.
The basic physical assumption is the dependence of all parameters on the volume scale factor of the Universe. Consequently,
the solution of the field equations can be represented in a parametric form. By supposing that the
volume-scalar field dependence is of exponential type, an exact solution corresponding to a dust filled Universe has been obtained.
The solution depends on two arbitrary numerical parameters, an integration constant and the coefficient $n$
relating the quintessence field to the volume scale factor of the Universe. The range of acceptable values of $n$ can be severely
restricted by imposing the conditions of accelerated expansion and that the energy density
of the $\phi $ field be of the same order of magnitude as the matter energy density.

In Fig. 1 we present the time variation of the energy density of the dust
matter and of the quintessence field, for different values of $n$. The
energy density of matter is practically insensitive to the value of $n$, its
time variation being mainly determined by the evolution of the volume scale
factor. For the chosen value of the integration constant $g$ there are two distinct periods in the evolution of the Universe.
During the first phase the matter energy density dominates the scalar field
energy density, $\rho _{m}>\rho _{\phi }$, but after a critical time $t_{c}$%
, the energy density of the quintessence field becomes larger than the
energy of the cold matter and the dynamics of the Universe is determined by
the scalar field. The variation of the deceleration parameter $q$,
represented in Fig. 2, shows a similar evolution, with decelerating
expansion during the matter domination phase and accelerated evolution for $%
t>t_{c}$. In the large time limit the deceleration parameter tends to $q=%
\frac{9}{2}n-1$, leading to an approximate range of  $q\in \left(
-1,-0.5\right) $. The variation of the pressure of the quintessence field,
presented in Fig. 3, shows that for all times $p_{\phi }<0$ and thus a negative pressure
drives the accelerate expansion of the Universe. But for larger values of $g$ the energy density of the
scalar field dominates the matter energy density from the early beginning of the cosmological evolution.

\begin{figure}
\epsfxsize=10cm
\centerline{\epsffile{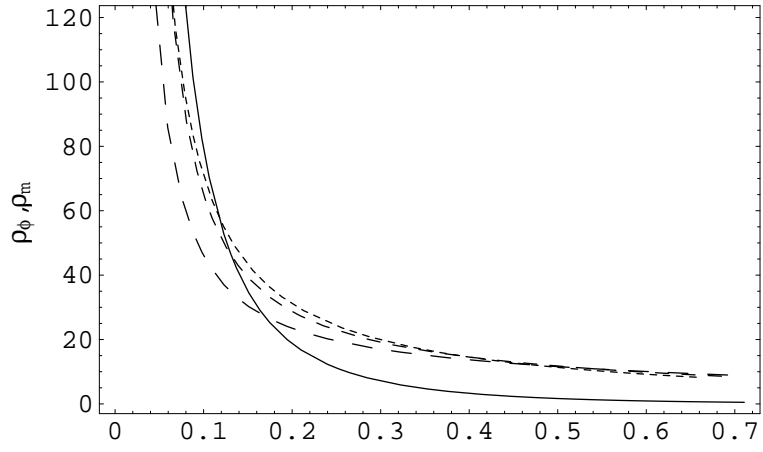}}
\caption{Variation as a function of the time $\tau =\sqrt{\rho _{0}}\left( t-t_{0}\right) $
of the energy density of the matter $\rho _{m}$ for $n=0.11$ (solid curve) and of the quintessence field $\rho _{\phi }$  for 
$n=0.11$ (dotted curve), $n=0.09$ (dashed curve) and $n=0.07$ (long dashed curve)
. The constant $g$ has been normalized by means of the condition $\frac{g}{\rho _{0}}=30$.}
\label{FIG1}
\end{figure}

\begin{figure}
\epsfxsize=10cm
\centerline{\epsffile{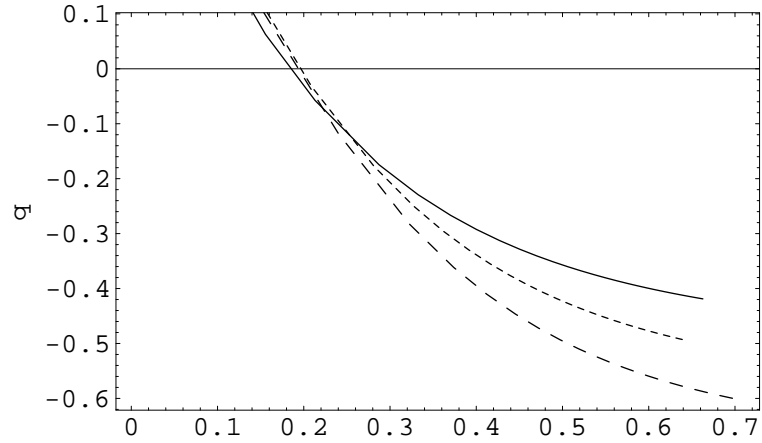}}
\caption{Variation as a function of the time $\tau =\sqrt{\rho _{0}}\left( t-t_{0}\right) $
of the deceleration parameter $q$ for $n=0.11$ (solid curve), for $n=0.09$ (dotted curve) and for $n=0.07$ (dashed curve).
The constant $g$ has been normalized by means of the condition $\frac{g}{\rho _{0}}=30$.}
\label{FIG2}
\end{figure}

\begin{figure}
\epsfxsize=10cm
\centerline{\epsffile{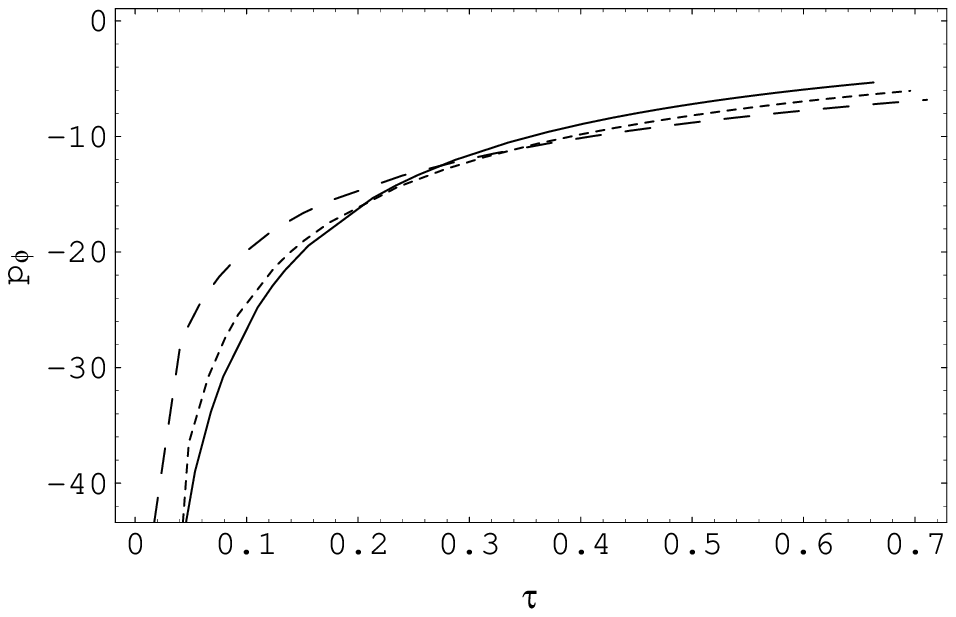}}
\caption{Variation as a function of the time $\tau =\sqrt{\rho _{0}}\left( t-t_{0}\right) $
of the pressure $p_{\phi }$ of the quintessence field for $n=0.11$ (solid curve), for $n=0.09$ (dotted curve) and for $n=0.07$ (dashed curve).
The constant $g$ has been normalized by means of the condition $\frac{g}{\rho _{0}}=30$.}
\label{FIG3}
\end{figure}

In this model the evolution of the Universe is generally insensitive to the
numerical values of the two constants $n$ and $g$ involved in the physical model.
For all times the energy densities of matter and scalar field are of same
order of magnitude. Therefore no fine-tunning of
parameters is necessary to explain the present state of the Universe and the coincidence
problem is also solved naturally, without imposing special initial conditions
for the quintessence field or its potential.

In the limit of large time, $V\rightarrow \infty $, we obtain $V\sim t^{2/3n}
$ and $a\sim t^{2/9n}$. In the same limit the other physical quantities
behave like
\begin{eqnarray}
\rho _{m}(t) &\sim &t^{-\frac{2}{3n}},\rho _{\phi }(t)\sim \frac{g}{3}t^{-2}+%
\frac{3\rho _{0}}{1-3n}t^{-\frac{2}{3n}},  \nonumber \\
U\left( t\right)  &\sim &\left( \frac{1}{3}-\frac{n}{2}\right) gt^{-2}+\frac{%
15n-4}{6(1-3n)}\rho _{0}t^{-\frac{2}{3n}}.
\end{eqnarray}

In the large time limit the deceleration parameter is a constant.

The gravitational field  equations with quintessence fields with arbitrary
self-interaction potentials are of a considerable complexity even for the
flat FRW geometry. Because of this mathematical complexity usually only very
simple models, with a power-law dependence of the scale factor and of the
scalar field, have been investigated. The more complicated models presented
in this paper can maybe provide a more complete description of the early and
late time behavior of our Universe. In particular these models lead to a
late time non-decelerating evolution of the Universe, with numerical values
of the deceleration parameter concordant to the astronomical observations
\cite{1}, \cite{2}. On the other hand the energy contribution of the quintessence
scalar field $\phi $ brings the total density parameter $\Omega $ very close
to the critical value $\Omega =1$.

\end{document}